# New Cybernetics and the Application of its Principles in Physics

**Oleg Kupervasser**

Transist Video LLC
119296 Skolkovo, Russia

Scientific Research Computer Center
Moscow State University
119992 Moscow, Russia

E-Mail: olegkup@yahoo.com

*We describe principles of new cybernetics and use these principles for resolution of basic physical paradoxes. It demonstrates universality of the principles of new cybernetics.*

## 1. Introduction

The key idea in cybernetics is feedback [**1**]. It is a closed cycle: the object affects itself through other objects. Within the scope of the new cybernetics [**2, 3, 4, 5**], this idea has been developed further - the cyberneticians (investigators and observers) and cybernetics (the studied/observed system) influence each other and thus create feedback. The new cybernetics is otherwise known as the cybernetics of cybernetics, or second-order cybernetics. Heinz von Foerster attributes the origin of second-order cybernetics to the attempts of classical cyberneticians to construct a model of the mind [**2**]: "a brain is required to write a theory of a brain. From this follows that a theory of the brain, that has any aspirations for completeness, has to account for the writing of this theory. And even more fascinating, the writer of this theory has to account for her or himself. Translated into the domain of cybernetics; the cybernetician, by entering his own domain, has to account for his or her own activity. Cybernetics then becomes cybernetics of cybernetics, or second-order cybernetics."

A set of basic features of the new cybernetics associated with this idea can be formulated:

- **A new type of feedback [2]:** New type of feedback is the connection of the observed system to itself by means of the observer. The new type of feedback also serves as the connection of the observer to him- or herself by means of the observed system. This connection is often small and can be neglected, but in many important cases, the connection cannot be ignored because of **the instability and randomness** of many real systems, which are discussed below.
- **Mutual influence and correlation of the observer and the observed system [2, 6]:** The inspected (observed) system cannot be viewed in isolation from the inspector (observer). There are interactions and correlations between them that often must be accounted for, in spite of their apparent smallness, because of **the instability and randomness** of many real systems, which are discussed below.
- **Relativity of the observers [6, 7]:** Different observers can see the observed system in different ways, but this does not lead to a contradiction.
- **Impossibility of complete self-description [6]:** The system cannot fully describe and predict (understand) itself. In fact, an attempt to perform a complete self-description of the system leads us to a contradiction. For example, we use ink to describe the system. However, this ink is also included in the system during the self-description process. That is, this ink also must be described. This can be performed with the use of different ink, but then this new ink also must be described. This process can be continued ad infinitum. However, an incomplete (partial) self-description is possible.
- **Complexity [8]:** Real systems consist of a large number of parts that interact with each other and the environment.
- **Instability and randomness [9]:** Many real systems are extremely sensitive to weak external interaction. This weakness leads to the importance of even the smallest interaction of the observer with the observed system.
- **Emergence [10]:** The whole often acquires new properties that cannot be predicted on the basis of even a complete knowledge of its parts and the interactions among them (the principal emergence). Indeed, we know that the observed system interacts with the observer. This interaction leads to the fact that the observed system be open and exposed to unpredictable external noise. Consequently, the complete system must include the observer. However, if the observer is included in the system, the latter becomes unpredictable because of **the impossibility of complete self-description**.
- **Integrity and relationship (correlation and interaction) [6, 10, 11, 12]:** Our world is not a collection of random bodies and events but rather a holistic set of cooperating (albeit sometimes very weak) objects that correlate with each other. That is, our world is more like an interconnected organism than a complex of random events (a holographic model of the universe or a holographic model of the brain, for example).
- **Scientific unpredictability (uncertainty) [6, 10]:** Many systems cannot be described in detail by using purely logical, scientific methods, such as mathematical modelling. Many systems, though, are unpredictable but feature probabilistic laws. Here, we discuss stronger unpredictability, which is not featured even by probability theory. Often, this impossibility of prediction is principal and not associated with the complexity of the system. This interaction leads to the fact that the observed system be open and exposed to unpredictable external noise. Consequently, the complete system must include the observer. However, if the observer is included in the system, the latter becomes unpredictable because of **the impossibility of complete self-description**.

- **Principal parameters of the system [6, 13]:** A **complex system often involves many par**ameters that describe it. However, in many cases, the dynamics of the system and its interface can be accurately described using only a small number of parameters. These are called the principal parameters of the system. The principal parameters constitute one of the methods of overcoming **the complexity** and **unpredictability** of the system. Examples of such principal parameters are the thermodynamic variables in physics and "features", which are the characteristics of multi-pixel objects used in the theory of pattern recognition.
- **Intuition (insight) [6. 14]:** In psychology, intuition occurs when a solution is reached very quickly and suddenly. What cannot be achieved by scientific methods can be achieved by means of intuition. By definition, intuition does not derive from science. Intuition is a method of overcoming the scientific **unpredictability** of the system *in practice*. In fact, the observer interacts with the outside world, is correlated with it, and acts as a part of it. This allows him or her to simply act on intuition (from some unexplained internal motivation) to achieve goals that cannot, in principle, be achieved by means of purely scientific methods.
- **Interdisciplinary approach [15]:** Very similar properties and phenomena occur in systems that belong to completely different areas of knowledge.

The main features of the new cybernetics can be illustrated with the help of examples from various areas of knowledge.

**Mathematics and logic:**

In mathematics, feedback is common and leads to the proof of the most fundamental theorems of impossibility. Here are some examples:

- Godel's incompleteness theorem [**16-17**]: Gödel introduced a single-correspondence function between a set of characters and the integers. This allowed him to formulate assertions about the properties of the axioms of a formal mathematical system as theorems regarding the arithmetic of integers. He could thus make a clear statement "I am lying", which refers to itself (**feedback**). Such a statement is mutually contradictory and can be neither true nor false. Thus, it can neither be proved nor disproved if the system of axioms is consistent. This leads to Gödel's incompleteness theorem: every self-consistent formal system of axioms, including the arithmetic, is principally incomplete.
- Gödel-Turing theorem: One can prove the impossibility of an algorithm that can be determined in a finite time whether a different algorithm will work indefinitely [**18**], based on the given input parameters, or will stop. An attempt to apply this algorithm to the input of itself (**feedback**) will lead to an inevitable contradiction, which proves its impossibility.
- Set theory [**19**]: Let us define the set of all sets that do not include themselves as subsets. Will such a set include itself (**feedback**)? Trying to answer this question inevitably leads to a contradiction. This means that, for the purpose of the consistency of the theory of sets, it is necessary to impose restrictions on the ability of certain aggregates to be sets.

**The neural theory of mind:**

To illustrate the principles of the new cybernetics and the proofs of the impossibility theorems, let us present a few examples of the neural theory of mind:

- **Is it possible to create an electronic mind that completely copies the work of neurons in the human brain and is equal to the human mind in terms of strength?** [**6, 20, 21, 22**] The answer to this question is as follows: not completely. Indeed, the work of neurons is determined by the internal state of the neurons, which, in its entirety, is equivalent to the full range of its constituent atoms. Next, the work of brain neurons depends on not only the neurons themselves but also the surrounding cells (glial cells). These cells depend on the operation of the whole organism. At the same time, the operation of the whole organism depends on the environment. That is, to simulate the brain in its entirety, it is necessary to simulate the entire world around it in its entirety to an approximation of atoms. This, in principle, cannot be done because of **the impossibility of complete self-description [6].** Of course, this does not mean that the operation of the mind cannot be modelled. In a certain approach, the operation of the mind can be modelled by introducing **the principal parameters** that describe neurons. However, it is highly unlikely that it can be done in its entirety, including the most subtle intuition. A little joke serves as an example. An apple fell on Newton's head, which led to the brilliant discovery of the law of gravity. Therefore, to create an electronic mind that can replicate this finding, we must simulate not only the mind of Newton but also the whole world around us, which led to the fall of that apple on his head at that very precise moment (**a new type of feedback** between the observer and the observed system)!
- **Is it possible to create a man and his brain that completely copies someone else's?** [**20**] No, it is not. The answer and its reasoning are the same as in the previous paragraph. There is no principal difference between exact physical copying and copying onto an electronic medium.
- **Is it possible to create a "code of the brain" that establishes single correspondence between the thoughts and the work of the neurons? In addition, is it possible to create a device that is thus capable of "reading minds"** [**6, 20, 21**]**?** In spite of the fact that modern science says that consciousness (or at least its observable manifestations) is completely determined by physical processes, the answer is the same: not completely. This is easily illustrated by **feedback [20].** Let us suppose that such a device is created and that it can read a person's thoughts (for example, define his or her anger). Moreover, imbued with full confidence in the device, the person can use it to judge his or her condition. However, what does the device show if it is used by the person to understand him- or herself? What happens? New feedback mechanisms of the brain appear, and a new "code of the brain" is formed, which includes not only neurons but also the mind-reading device. Moreover, such feedback can occur naturally. For example, a person may assess his or her wrath from the sweat on his or her forehead or the fright of nearby people. They play the role of the mind-reading device. Thus, the code of the brain in its entirety depends on not only neurons but also other cells of the body and even the environment. Furthermore, it can change dynamically over time. This does not mean that the "code of the brain" is completely inscrutable. The **principal parameters** of neurons can describe it. However, there are natural limits to the completeness of such a description, which are specified above.

**Computer science:**

- **Is it possible to create a "Chinese Room" in reality, at least in principle?** [**6, 20, 21**] Not in its entirety. Just as we have shown above, the full set of rules for the human person in the "Chinese Room" is equal to the size of the universe and must include that person, which cannot happen because **the impossibility of complete self-description**.
- **Can a computer behave unpredictably** [**6**]**?** Yes, it can. An example is the quantum computer (or its classical counterpart: an unstable analogue computer [**23-25**]). Indeed, the work of such a computer can only be assessed by the person who initiated it. Any outsider who was not present at its start-up will only violate its operation when trying to understand how it works. Such a system is unpredictable for him.
- **Can an intellect that is comparable to that of a human being work only on the basis of scientific logic and not possess intuition?** [**6, 26**] No. We have said above that the presence of **unpredictable** systems limits the scope of purely scientific knowledge of the world.
- **Can a computer possess intuition and guess better than the genetic algorithm?** [**6, 26**] Yes. To model intuition, computers often use a random number generator. However, such a model is primitive [**27**]. For a more accurate modelling of intuition, it is necessary to allow the environment to influence this generator. Then, correlations appear between the environment and the generator. Introducing such correlations may allow the computer to better "guess" the properties of the environment than a simple random generator does in the genetic algorithm. It should be noted that this intuition bears a high price: the computer itself becomes an unpredictable system, which can lead, for example, to a robot revolution.
- **Can the intuition of an artificial intelligence be fully equivalent to that of a human being in terms of efficiency?** [**6**] If humans were able to clearly formalise their goals and desires, the answer would be yes. However, many desires of people cannot be formalised and are not clear even to them. As we have written above, it is impossible in principle to simulate the human in its entirety. Therefore, a perfect machine intuition cannot be created either. Only human beings themselves know best what they need.
- **Is it possible to create an artificial intelligence that is equivalent to that of men in terms of efficiency**? [**3, 6**] No. Because human intuition cannot be copied in its entirety, this means that human beings will always perform certain tasks better than the machine can because they possess a better understanding of their desires. Moreover, a possible uprising of robots is the best example of such an inadequacy of the artificial intelligence because it is unable to solve the problems that require solution for the person as such a person or better.
- **How is it possible to create an artificial intelligence comparable in strength to a human for the solution of formalisable problems (problems with a precise description and a precise purpose)?** It seems that such an intelligent system should be created similarly to the evolution and education of a human person but at a more accelerated rate. It should be a training and self-learning system. Most likely, the most suitable model will be similar to a neural network. Even existing heuristic algorithms should be transformed into neural networks. The intelligent system should be created from simple forms into complex forms: it is necessary to begin from imitating the behaviour of the simplest insects, which have the most primitive nervous systems. Then, it will be necessary to complicate the task gradually, creating more and more complex robots, using previous results for subsequent stages. It will be necessary to use something like natural selection. This is the genetic algorithm. If we wish to model intuition (for the creation of such an intelligent system or for the solution of non-formalisable problems), it will be necessary to use a generator of random numbers that is correlated with the investigated environment.

**Economic, political, psychological, and sociological sciences:**
Here, the importance of the above principles is evident [**28**]. The influence of the observer on the observed system is very large. Any "theory" that captures the minds of many becomes a part of the inspected system itself. For example, the winning theory in the stock market is impossible; if it were true, all would have to win, but that is impossible. Therefore, these sciences cannot be completely considered to be science. Intuition has always played a huge role here, a role that is much more significant than in other sciences.

**Physics:**
Physics would seem to be the most exact science, in which the impact of the observer is immaterial, as opposed to the economic, political, psychological, and sociological sciences [**28**]. How can the observer influence the results of experiments with elementary particles in an accelerator? However, for many real physical systems, the observer does indeed influence the system; these systems possess the properties of instability and randomness. This leads to the resolution of many paradoxes in physics. The remainder of this work is dedicated to such situations in physics. We will give only one example here.
**Why is the beginning of the universe characterised by low entropy? Why cannot our universe be the result of a gigantic fluctuation from the thermodynamic equilibrium with maximum entropy?** The answer is simple: in fact, it can be. Moreover, this is more likely under the laws of statistical physics. However, experimentally determining which of the above two theories of the origin of the world is true is impossible in principle because of **the impossibility of complete self-description** and self-observation. Feynman [**29**] argues that in the case of a giant fluctuation uprising, there would have to be an "ocean" of thermodynamic equilibrium around us, and he sees a contradiction here. However, there is none. The observer may simply be unable to observe this "ocean". After all, he is a part of this fluctuation and interacts with this world. He cannot break out of its limits, as his uncomfortable environmental equilibrium will be destroyed just by his trying to observe the surrounding "ocean". Thus, the low-entropy theory of the initial universe is simply a convenient point of view rather than an established scientific fact. We can see only a small part of the observable universe, and yet, we try to assess the entirety of the universe on the basis of nothing but a small part of it, which is a very error-prone approach.

## 2. Using principles of new cybernetics for resolution of physical paradoxes

A recent letter by Maccone [**30**] presents a solution based on the existing laws of quantum mechanics to the arrow-of-time dilemma. He argues that all phenomena in which the entropy decreases must not leave any information (in the observer's memory) of their having occurred because the observer is a part of the whole system. Maccone concludes that quantum mechanics is necessary to his argument, which he believes does not otherwise work in classical mechanics. This part consists of four sub-parts. We discuss the basic problems in the first part. This Comment and the previously published Comment by Jennings and Rudolph describes flaws in Maccone's arguments. However, the main argument (erasure of the observer's memory), which was previously formulated in our work and was repeated by Maccone, is correct under the conditions described in this Comment. Moreover, this argument can be used to resolve a reduction paradox (the Schrödinger's Cat paradox) in quantum mechanics. This use is demonstrated in the second part. In the third part, the synchronisation (decoherence) of time arrows is discussed. In the fourth part, the synchronisation (decoherence) of time arrows in quantum gravity is considered.

### 2.1. Macroscopic entropy, entropy increase law and memory erasure argument

#### 2.1.1. The relevance of memory erasure argument for classical mechanics.

A recent Letter by Maccone [**30**] presents a solution based on the existing laws of quantum mechanics to the arrow-of-time dilemma. He argues that all phenomena in which the entropy decreases must not leave any information (in the observer's memory) of their having occurred because the observer is a part of the whole system. He concludes that quantum mechanics (QM) is necessary to his argument, which, he believes, does not otherwise work in classical mechanics (CM). Papers [**31-33**] have clearly shown that the same arguments hold true for *both* quantum and classical mechanics. Thought experiments of both the Loschmidt (time reversal paradox) and Poincare (recurrence theorem) type are used to illustrate the arrow-of-time dilemma in the latter papers whereas Maccone uses only Loschmidt's experiment; however, he then gives a mathematical proof for the general case with entropy decrease.

The arguments to resolve both paradoxes in classical mechanics are as follows. At least in principle, CM allows exclusion of any effect from the observer on the observed system. However, most real systems are *chaotic* – a weak perturbation may lead to an exponential divergence of trajectories; also, there is always a non-negligible interaction between the observer and the observed system. Let us take the simple example of a gas expanding from a small region of space into a large volume. In this entropy-increasing process, the time evolution of the macroscopic parameters is stable to small external perturbations. After some time, if all the velocities are reversed, the gas will return to the initial small volume; this is true in the absence of any perturbation. This entropy-decreasing process is clearly unstable, and a small external perturbation would trigger continuous entropy growth. Thus, entropy-increasing processes are stable, but entropy-decreasing processes are unstable. A more rigorous theory has been developed for the general case [**31-33**]. The natural consequence of this theory is that the time arrows (whose direction is defined by the entropy growth) of both the observer and the observed system are synchronised due to the inevitable non-negligible interaction between them. Both the observer and observed system can only return to the initial state together (as the whole system) in both the Loschmidt and Poincare paradoxes; thus, the observer's memory is erased in the end. Approaching the end point, the observer's time arrow is opposite to the coordinate (absolute) time arrow, and entropy growth is observed in the entire system and in both parts of the system although the entropy decreases in the absolute time.

#### 2.1.2. The entropy increase law is FAPP law in both CM and QM

It is important to remark that the unobservability of the entropy decrease is correct only for certain *practical cases* of perturbative QM measurement experiments. For an ideal nonperturbative observation and a thermodynamically correct definition of the system entropy, the entropy decrease can *in principle* be observed in the framework of QM.

Let us first define a nonperturbative observation [**31-33**] in QM. Suppose we have some QM system in a known initial state. This initial state can be either the result of some preparation (e.g., an atom comes to the ground electronic state in vacuum after a long time) or the result of a measurement experiment (a QM system after measurement can have a well-defined state corresponding to the eigenfunction of the measured variable). We can predict further evolution of the initial wave function. Therefore, *in principle*, we can make further measurements by choosing the measured variables such that one of the eigenfunctions of the current measured variable is a current wave function of the observed system. Such a measurement process can allow for continuous observation without any perturbation of the observed quantum system. This nonperturbative observation can be easily generalised for the case of a known, *mixed* initial state.

For example, let us consider a quantum computer (QC). It has some well-defined initial state. An observer that knows this initial state can *in principle* make a nonperturbative observation of any intermediate state of the QC. However, an observer that does not know the initial state cannot make such observation because he cannot predict the intermediate state of the QC.

We can conclude that the entropy increase law is for all practical purposes (*FAPP)* law. It is correct for *perturbative* observations of macroscopic quantum systems and classical macroscopic chaotic systems due to the erasure of the observer's memory. Such small perturbations exist in any real case. However, in a general case, this is not correct.

#### 2.1.3. Correct definition of thermodynamic entropy in the Loschmidt paradox

The purpose of Maccone's study was to resolve the Loschmidt paradox between the second law of thermodynamics and the reversibility of motion. Therefore, the thermodynamically correct definition of entropy, which is actually used in the formulation of the second law, must be chosen. Let us give such a definition for the entropy. Two different definitions of the entropy can be made: *macroscopic* and *ensemble* entropies [**31-33**]. The macroscopic entropy is the entropy calculated from the macroscopic parameters for *all* of the microstates that have these parameters whereas the ensemble entropy is calculated for *some* set of microstates that evolved over time from the initial state. Lastly, a standard formula for the entropy (von Neumann or classical) over the obtained distribution must be used. The second law of thermodynamics law (regarding the increase in entropy) uses the macroscopic definition of entropy.

Maccone defines the system entropy to be the sum of the ensemble entropies of the observer and the observed system ($S(A \text{ and } C) \equiv S(\rho_A) + S(\rho_C)$). For a perturbative QM observation of the macroscopic system, his definition of entropy is equivalent to that of the macroscopic entropy because both the observer and the observed system are in mixed states and correlate through microscopic variables. The classical analogues that Maccone uses to prove the necessity of QM are wrong. In his mutual entropy formula $S(A{:}C) \equiv S(\rho_A) + S(\rho_C) - S(\rho_{AC})$, the *macroscopic* entropy of the subsystems should have been used upon the transition from QM to CM, not the *ensemble* entropies. Contrary to CM, in QM both the macroscopic and ensemble entropies have the same numerical value for perturbative QM observations, although the definitions of the entropies differ. For classical macroscopic chaotic systems (the observer and the observed system) and for non-negligible interaction between the observer and observed system, the *ensemble* entropies can also be used. However, the *initial states* of the observer and observed system must be calculated from macroscopic parameters for *all* the microstates that have these parameters.

However, generally, in classical and quantum mechanics (e.g., for nonperturbative observations), this definition is not a correct definition of the thermodynamic entropy of the system. Indeed, let us consider a simple example of gas expanding from a small region of space into a large volume. This process is a macroscopic entropy-increasing process. We must use the thermodynamically correct, *macroscopic* entropy of the ideal gas: $S=kNlnV+const$ $(T=const)$. Based on the Poincare return theorem, the gas will be very close to the initial small volume after a long time; this result is true in the absence of any perturbation. This process is a macroscopic entropy-decreasing process. In contrast to the macroscopic entropy, the ensemble entropy of the gas does not change during this evolution. Suppose we know the initial quantum state of this gas; *in principle*, we can make the nonperturbative observation described above. We therefore will be able to observe both the initial entropy increase and the final entropy decrease. This result contradicts the primary conclusion of Maccone's study. However, Maccone's considerations and conclusions are correct for the *practical case* of a *perturbative QM* observation of a macroscopic system. In this case, we used a *fixed* set of macroscopic variables for the observation. This set does not depend on the initial state (in contradiction to the nonperturbative observation described above).

We find no flaw in Maccone's *entropic considerations* within QM for perturbative observations of macroscopic systems. In contrast, D. Jennings and T. Rudolph [**34**] objected to this definition of entropy. However, the objection of D. Jennings and T. Rudolph is not relevant here [**35**] because we consider macroscopic systems. However, the examples of D. Jennings and T. Rudolph correspond to microscopic systems.

### 2.2. Schrodinger's Cat paradox and spontaneous reduction

The complete violation of the wave superposition principle (i.e., the full vanishing of interference) and reduction of the wave function would occur only during the interaction of a quantum system with an ideal macroscopic object or device. The ideal macroscopic object either has infinite volume or consists of an infinite number of particles. Such an ideal macroscopic object can be consistently described both by quantum and by classical mechanics.

Furthermore, similarly to the classical case, we consider only systems with finite volume and a finite number of particles (unless the other is assumed). Such devices or objects can be considered to be only approximately macroscopic.

Nevertheless, a real experiment shows that even for such non-ideal macroscopic objects, the destruction of superposition and the correspondent wave function reduction may occur. We will define such a reduction for imperfect macroscopic objects as *spontaneous reduction*. Spontaneous reduction leads to paradoxes, which force one to doubt the completeness of quantum mechanics despite its tremendous successes. We will reduce the most impressive paradox from this series – the Schrodinger's Cat paradox.

Schrodinger's Cat is a thought experiment that clarifies the principle of superposition and the reduction of wave functions. A Cat is placed in a box. In addition to the Cat, there is a capsule with poisonous gas (or a bomb) in the box; this capsule (or bomb) can blow up with 50 per cent probability due to the radioactive decay of a plutonium atom or a quantum of casually illuminated light. After some time, the box is opened and one learns whether the cat is alive or not.

Until the box is opened (if the measurement is not performed), the cat remains in a very strange superposition of two states: "alive" and "dead". For macro-objects, such a situation appears very mysterious (In contrast, for quantum particles, the superposition of two different states is very natural). Nevertheless, no basic prohibition of quantum superposition for macrostates exists.

The reduction of these states upon the opening of the box by an external observer does not lead to any inconsistency with quantum mechanics. This reduction is easily explained through the interaction of the external observer with the Cat during the measurement of the Cat's state.

Nevertheless, a paradox arises for the closed box if the observer is the Cat itself. Indeed, the Cat possesses consciousness and is capable of observing both itself and the environment. Upon introspection, the Cat cannot be simultaneously alive and dead but is in just one of these two states. Experience shows that any conscious creature feels itself to be either alive or dead. Both such situations do not exist simultaneously. Therefore, the spontaneous reduction to two possible states (alive and dead) truly occurs. The Cat, even with all the contents of the box, is not an *ideal* macroscopic object. Therefore, such an observable and nonreversible spontaneous reduction contradicts reversible Schrodinger quantum dynamics. In the current case, this contradiction cannot be explained by some external influence because the system is isolated.

Does this system actually contradict Schrodinger quantum dynamics? When is the system macroscopic sufficient to provide the possibility for spontaneous reduction? Is it necessary for such a nearly macroscopic system to have consciousness like a Cat?

The multi-world interpretation as such does not explain the Schrodinger's Cat paradox; the interpretation only reformulates and conceals the paradox. Indeed, the Cat observes only one of the existing worlds. However, the results of further measurements depend on the correlations between the worlds. Nevertheless, neither these worlds nor these correlations are observed. "Parallel worlds" that we know nothing about can always exist. However, these worlds can truly affect the results of some future experiment of ours. That is, knowledge of the current state only (in our "world") and of the laws of quantum mechanics does not even allow us to predict the future probabilistically! However, quantum mechanics was developed for such predictions! Solely on the basis of spontaneous reduction that destroys quantum correlations between worlds, we can predict the future with knowledge of only the current (and actually observed) states of our "world". The paradox of Schrodinger's Cat returns and has only changed its shape.

Remember that the paradox of Schrodinger's Cat consists of the inconsistency between the spontaneous reduction observed by the Cat and the Schrodinger evolution that forbids such reduction. To correctly understand the paradox of Schrodinger's Cat, it is necessary to consider the paradox from the perspective of two observers: the external observer-experimenter and the Cat (i.e., *introspection*).

For the external observer-experimenter, the paradox does not arise. If the experimenter attempts to discern whether the Cat is alive or not, the experimenter inevitably influences the observable system (in agreement with quantum mechanics) and leads to the reduction. The system is not isolated and hence, cannot be described by the Schrodinger equation. The reducing role of the observer can also be played by the surrounding medium. This situation is defined as decoherence. Here, the role of the observer is more natural and is reduced to registration of the decoherence. In both cases, there is entangling of measured system with the environment or the observer, i.e., there are correlations of the measured system with the environment or the observer.

What happens if we consider a closed complete physical system that includes the observer, observed system and environment? This is the case with the Cat's introspection. The system includes the Cat and his box environment. It should be noted *that full introspection (full in the sense of quantum mechanics) and full verification of the laws of quantum mechanics are impossible in the isolated system that includes the observer*. Indeed, in principle, we can measure and analyse the state of an external system precisely. However, if we include ourselves in the consideration, there are natural restrictions. These restrictions are related to the possibility of retaining memories and analysing states of molecules with the molecules themselves. Such an assumption leads to inconsistencies. Therefore, the possibility of finding an experimental inconsistency between Schrodinger evolution and spontaneous reduction through introspection in an isolated system is also restricted.

Nevertheless, let us attempt to find some mental experiments that lead to inconsistency between Schrodinger evolution and spontaneous reduction.

1) The first example is related to the reversibility of quantum evolution. Suppose we introduce a Hamiltonian capable of reversing quantum evolution in the Cat-box system**.** Practically speaking, this process is nearly impossible; however, theoretically, no problem exists. If spontaneous reduction occurs, the process would be nonreversible. If spontaneous reduction is not present, the Cat-box system will return to an initial pure state. However, only an external observer can make such verification. The Cat cannot make it by introspection because the Cat's memory will be erased upon restoration of the initial state. From the perspective of the external observer, no paradox exists because he does not observe the spontaneous reduction that truly can lead to a paradox.
2) The second example is related to the necessity of Poincare's return of the quantum system to an initial state. Suppose the initial state was pure. If the Cat has introspection and if spontaneous reduction truly exists, it leads to a mixed state. Then return would be impossible - the mixed state cannot transfer to a pure state through the Schrodinger equation. Thus, if the Cat has fixed return, the situation is inconsistent with spontaneous reduction. However, the Cat cannot fix return (in the case of quantum mechanics fidelity) because return will erase the Cat's memory. Therefore, there is no paradox. The exterior observer actually can observe this return by measuring an initial and final state of this system. However, no paradox exists there either because the observer does not observe any spontaneous reduction that actually can lead to a paradox. It is worthwhile to note that the inconsistency between spontaneous reduction and Schrodinger evolution can be experimentally observable only if memory of the spontaneous reduction is retained by the observer and if this memory is not erased or damaged. No experiments described above are covered by this requirement. Thus, these examples clearly show that, although spontaneous reduction actually can lead to violation of Schrodinger evolution, this violation is not observed experimentally (with fidelity to quantum mechanics).
3) The third example follows: Quantum mechanics gives superposition of a live and dead Cat in a box. Theoretically, an exterior observer can always *precisely* measure this superposition if this superposition is one of the measurement eigenfunctions. Such a measurement would not destroy the superposition, in contrast to the case in which the live and dead Cat are eigenfunctions of the measurement. Having informed the Cat about the result of the measurement, we will introduce inconsistency with spontaneous reduction observed by the Cat. Such an argument has a double error. *At first*, this experiment is used for *verification* of the Cat's spontaneous reduction of existence when the observer is the Cat itself. The external observer does not influence the Cat's memory only if spontaneous reduction is not present and the Cat's state is a superposition of live and dead states. However, the observer does influence and can destroy the Cat's memory if spontaneous reduction occurs. Therefore, such an experiment cannot legitimately verify the existence of spontaneous reduction in the *past*. *Secondly*, the data transmitted to the Cat is retained in his memory. Thus, this transmission changes both the state and all further evolution of the Cat; i.e., the system cannot be considered to be isolated after the measurement. Therefore, no contradiction with the *future* exists.

The external observer does not observe spontaneous reduction and hence, does not observe the paradox. Thus, from the perspective of the external observer, verification with the aid of continuous nonperturbative observation, described in term 3, is possible and is legitimate. This verification does not influence the external observer's memory. Moreover, such verification, which does not interrupt the evolution of the observable system, allows for the measurement of not only the initial and final states of the system but also all of the intermediate states. That is, this verification implements continuous, non-perturbative observation!

It should be noted that the external observer can only theoretically observe the superposition of alive and dead Cat. Practically speaking, this observation is nearly impossible. In contrast, for small quantum systems, superposition is very observable. This difference results in the fact that quantum mechanics is generally considered to be the theory of small systems. However, for small macroscopic (mesoscopic) objects observations of superposition are also possible. A set of particles at low temperature or the states of some photons [**36**] are examples. We make an important remark**:** recently, very interesting papers were published toward the construction of mesoscopic "synergetic" systems, which are most likely similar to living organisms [**30**], [**37-39**]. It must be mentioned that the construction of such models is a problem of physics and mathematics, not philosophy.

### 2.3. Synchronisation/decoherence of time arrows.

The follow question can arise. Let us assume that some process exists in which the entropy decreases. For definiteness, let us take this process to be the spontaneous reconstruction of a house (which was previously destroyed in an earthquake).

Let us also take the simple example of gas expanding from a small region of space into a large volume. If, after some time, all the velocities are reversed, the gas will return to the initial small volume.

If we use a camera to take a series of snapshots recording different stages of the spontaneous house construction (or gas shrinkage), we expect that the camera will record this spontaneous process. Why will the camera not be able to record it? What precisely will prevent the camera from recording these snapshots?

The answers to these questions are as follows: even a very small interaction between the camera and the observed system destroys the process of the inverse entropy decrease and results in the synchronisation of the direction of the time arrows of the observer and the observed system. (The direction of a time arrow is defined to be the direction of the entropy increase.) This very small interaction occurs because light illuminates the observed object and is reflected by the camera (and because light illuminates the camera). In the absence of the camera, the environment can act as the observer by being illuminated by and reflecting the light. (Any process without an observer is nonsensical. The observer must appear at some stage of the process; however, the influence of the observer is much smaller than the environmental influence). External noise (interaction) from the observer/the environment destroys correlation between molecules of the observed system. This noise prevents the inverse process with the entropy decrease. In quantum mechanics, such a process is defined as "decoherence". The house reconstruction (or gas shrinkage) will be stopped, i.e., the house will not actually be reconstructed/(the gas will not shrink). In contrast, processes in which the entropy increases are stable.

The following example is from Maccone's study [**30**]: "However, an observer is macroscopic by definition, and all remotely interacting macroscopic systems become correlated very rapidly (e.g., Borel famously calculated that moving a gram of material on the star Sirius by 1 m can influence the trajectories of the particles in a gas on earth on a time scale of s [**40**])"

Nevertheless, it is not a problem to reverse both the observer (the camera) and the observed system. From the Poincare return theorem for a closed system (which includes both the observer and the observed system), this return must occur automatically after a very long time. However, the memory erasure of the observer prevents this process from being registered.

The majority of real systems are *chaotic* – a weak perturbation may lead to an exponential divergence of trajectories, and there is also always a non-negligible interaction between the observed system and the observer/environment. However, *in principle*, in both quantum mechanics and classical mechanics, we can make nonperturbative observations of the entropy decrease process. A good example of such a mesoscopic device is a quantum computer: no entropy increase law exists for such a system. This device is very well isolated from the environment and the observer. However, *in practice*, nonperturbative observation is nearly impossible for macroscopic systems. We can conclude that the entropy increase law is *FAPP law*.

It should be mentioned that decoherence (synchronisation of time arrows and "entangling") and relaxation (during relaxation, a system achieves equilibrium) are absolutely different processes. During relaxation, macroscopic variables (entropy, temperature, and pressure) change greatly to their equilibrium values, and the invisible microscopic correlations between the parts of the system increase. During decoherence, the macroscopic variables (entropy, temperature, and pressure) are nearly constant. Invisible microscopic correlations inside of the subsystems (environment, observer, and observed system) are largely destroyed; however, new correlations appear between the subsystems. This process is named "entanglement" in quantum mechanics. During this process, synchronisation of the time arrows also occurs. The relaxation time is much longer than the decoherence time.

Let us consider the synchronisation of time arrows for two systems that are non-interacting (before some initial moment). It should be mentioned that this description is made in an absolute (coordinate) system. However, both systems also have their own initially opposite time arrows, which are defined by the direction of entropy growth in each system.

This description means that there exist two non-interacting systems such that in one system time flows (i.e., entropy increases) in one direction whereas in the other system time flows in another (opposite) direction. However, if the systems come into an interaction with each other, then one system (the "stronger" one) will drag the other ("weaker") system to flow in his ("stronger") direction, so eventually, they will both have time flowing in the same direction.

"To be stronger"-what does this mean, exactly? Is this strength something that increases with the number of degrees of freedom of the system? This supposition is not correct except for small fluctuations. "Stronger" or "weaker" does not appreciably depend on the number of degrees of freedom of the systems. The interaction described above is asymmetric in the absolute (coordinate) time. For the first system, the interaction appears in its *future* after the initial moment (At the initial moment the systems have opposite time arrows) based on the time for this system. For the second system, the interaction was in its *past* based on the time of this system. Therefore, the situation is *not symmetric in time*, and the first system is always "stronger". This occurs due to the instability of processes that decrease entropy and the stability of the processes that increase entropy, as described above.

Indeed, let us consider again two initially isolated vessels of gas. In the first, the gas expands (the entropy increases). In the second, the gas shrinks (the entropy decreases).

In the first vessel, the gas expends from a small volume in the centre of a vessel. The velocities of the molecules are directed from the centre of the vessel to its boundary. It is physically clear that a small perturbation of the velocities cannot stop the expansion of gas. Indeed, after a random small perturbation, the velocities will continue to be directed from the centre of the vessel to its boundary. Noise can even increase the expansion. Therefore, the expansion process is stable.

In the second vessel, the gas shrinks from the full volume of the vessel to its centre. The velocities of all the molecules are directed toward the centre of the vessel. It is physically clear that a small random perturbation of the velocities can easily stop the shrinkage of the gas. Indeed, even after a small perturbation, the velocities will not be directed toward the centre of vessel. Thus, the shrinking process is stopped. We can therefore conclude that the shrinking process is unstable. This shrinking process can be obtained by reversing the expansion of the gas. If we reverse the velocities of the molecules of the expanding gas *before* the collisions of the molecules with each other and the vessel boundary, this instability is linear and is not strong. However, if the reversal occurs *after* these collisions, this instability is exponential and is much stronger.

Both time directions have equal roles. However, a small random noisy interaction breaks this symmetry for the two systems described above due to the instability of the entropy decrease processes. Time symmetry exists only for the *full* system that includes the two subsystems defined above. However, the time arrows of the interacting subsystems must become the same.

Instead of an interaction with infinite time $[0, +\infty]$ we can consider an interaction with a large finite time T: [0, T]. Let us choose this time T to be much smaller than the Poincare return time. Thus, in the first system we have an interaction during [0, T] based on its own time, and in the second system, the interaction is during [-T, 0] based on its own time (the "0" time moment for the first system corresponds to the "–T" time moment in the second system). Can we still apply our argument? Instead of the asymmetry of the forces, in this case, we obtain an asymmetry of the initial conditions. At the initial moment 0 for the first coordinate system [0, T], the two vessels have different time arrows. However, at the initial moment -T for the second coordinate system [-T, 0], the two vessels have the same time arrows in a negative direction.

Only if T is exactly equal to the Poincare return time will the situation indeed be symmetric. For such a situation, the two time arrows are also different at the moment T, but each arrow is opposite its initial direction at time 0. Again, the "stronger" system has the interacting forces in its future with respect to its own time arrow.

This theory can explain how entropy growth occurs in the same direction in all parts of the Universe. However, this theory cannot explain the low entropy, initial condition of the Universe. This condition is most likely a result of the anthropic principle [**41**].

## 2.4. The law of entropy increase and "synchronisation of time arrows"/decoherence in gravitational theory.

### 2.4.1. Black Holes

In Einstein's theory of general relativity, similarly to classical mechanics, motion is reversible. However, an important difference also exists between general relativity and classical mechanics. General relativity is an ambiguous theory. Indeed, in general relativity, two different initial states can give infinitesimally close states after a *finite* time interval. This situation occurs, for example, during the formation of a black hole because of a collapse. Let us consider the inverse process, which describes a white hole. In this process, initial states that are infinitesimally close after a *finite* time interval can give different final states. Thus, an observer/environment can considerably affect the evolution of the state during the *finite* time interval even if the observer/environment has an infinitesimally weak interaction with the white hole. We must mention that in contradiction to CM and QM, in gravitational theory, an arbitrarily small but finite interaction *always* exists. Gravity forces always exist between two arbitrary objects with nonzero masses.

Because of these two properties, the law of entropy increase is an exact law but not FAPP in general relativity theory. Therefore, entropy becomes a fundamental concept. Indeed, there is such fundamental concept as the entropy of a black hole. In addition, it is possible to explain the existence of this entropy by the perturbation created by the observer. Unlike in classical mechanics, this perturbation may now even be infinitesimally weak. During formation of a black hole, entropy increases.

Time reversal leads to the appearance of a white hole and an entropy decrease. In reality, a white hole cannot exist because of the entropy decrease. An entropy decrease is prohibited in general relativity for the same reason that it is prohibited in classical mechanics. This instability of the entropy decreasing processes is much stronger in general relativity than it is in classical mechanics. This instability results in the synchronisation of the time arrows of the white hole and of the observer/environment. The direction of the time arrow of the white hole changes to coincide with the time arrow of the observer/environment. The white hole transforms into a black hole.

Here is also the well-known black hole information paradox [**42**]**:** information (which in classical and quantum mechanics is conserved) disappears in a black hole forever. It would appear that there is no problem; the information is most likely stored inside of the black hole in some form. However, chaotic Hawking radiation makes this process of information loss explicit; the black hole evaporates, but the information is not recovered.

Hawking radiation concerns semiclassical gravitation. However, the paradox can also be formulated within the framework of the theory of general relativity. A spherical black hole can be reversed into a white hole at some moment. (This process appears impossible, but a physically similar situation with “wormholes” connecting black and white holes in different universes is considered in [**43**]). Thus, the process is converted in time. Nevertheless, information cannot be recovered due to the ambiguity (the infinitely strong instability) in the evolution of the white hole.

Usually only two solutions for this problem are considered. Either the information truly disappears or because of interior correlations of the Hawking radiation (or the exact reversal of the black hole processes after its transmutation to a white hole) the information is conserved. However, most likely, a third solution is true. Due to the inevitable influence of the observer/environment it is impossible to distinguish these two situations experimentally! However, if it is impossible to confirm this result experimentally, it is not a scientific subject.

Both in the theory of general relativity and for semiclassical gravitation the paradox can be resolved by the influence of the observer/environment. Indeed, let us suppose that Hawking radiation is correlated, not chaotic (or the white hole would be the exact inverse of the black hole). Thus, the infinitesimal influence of the observer/environment leads to the inevitable losses of these correlations (and the corresponding information) during the finite time interval. It is senseless to include the observer in the described system: complete self-description and introspection is impossible. In such a situation, the law of conservation of information cannot be confirmed experimentally even if it is actually correct.

Currently, we have no general theory of quantum gravitation. However, for a special case of a 5-dimensional anti-de-Sitter space, many scientists consider this paradox be resolved. Information is supposed to be conserved because of a hypothesis regarding AdS/CFT dualities; i.e., the hypothesis that quantum gravitation in the 5-dimensional anti-de-Sitter space (that is with a negative cosmological term) is mathematically equivalent to a conformal field theory regarding a 4-surface of this world. This hypothesis was confirmed for some special cases but is not yet proved for the general case. Suppose that this hypothesis is actually true. At first glance, this hypothesis automatically solves the information problem. Conformal field theory is unitary. If conformal field theory is actually dual to quantum gravitation, then the corresponding quantum gravitational theory is unitary as well. Therefore, in this case, information is not lost. However, we suppose that this hypothesis is not correct. The process of the formation of a black hole and its subsequent evaporation occurs on *all surfaces* of the anti-de-Sitter space (described by conformal quantum theory). This process also includes the observer/environment. However, the observer cannot precisely know the initial state and cannot analyse the behaviour of the system to verify unitarity because he is a part of this system! Hence, the observer’s influence on the system cannot be neglected. Thus, experimental verification of the information paradox again becomes impossible!

### 2.4.2. Wormholes

Let us consider from the perspective of the entropy increase law a paradoxical object in the general relativity theory: a wormhole [**44**]**.** We will consider a Morris-Thorne wormhole [**45**]**.** Through a very simple procedure (we place one of the mouths of the wormhole on a spaceship; then the spaceship moves with relativistic velocity over a closed loop and returns the mouth to its initial location), a wormhole traversing space can be transformed into a wormhole traversing time. After this transformation, the wormhole can be used as a time machine and leads to the well-known grandfather paradox. How can this paradox be resolved?

For macroscopic wormholes, the solution can be found through the entropy increase law. The realisation of this law is ensured by the instability of entropy decreasing processes, and this instability results in the synchronisation of time arrows.

Indeed, a wormhole traversing space does not lead to a paradox. If an object enters one mouth at some moment in time, then it exits the other mouth at some later moment in time. Thus, the object travels from an initial, high-order, low-entropy environment to a future, low-order, high-entropy environment. During the trip along the wormhole, the entropy of the object also increases. Thus, the directions of the time arrows of the object and the environment are the same. The same conclusions are correct for travelling from the past to the future through a wormhole that traverses time.

However, for travel from the future to the past, the directions of the time arrows of the object and the environment are opposite. Indeed, the object travels from the initial, low-order, high-entropy environment to the high-order, low-entropy environment. However, the entropy of the object increases! As previously described, such a process is unstable and will be prevented or will be forcibly converted through a synchronisation process of the time arrows.

The initial synchronisation of a wormhole with its environment must occur when the moving mouth of the wormhole returns to its initial state during its creation.

How does the environment appear inside of the wormhole? The massive ends of the wormhole radiate. This thermodynamic radiation appears inside of the wormhole. This radiation is the environment of a traveller inside of the wormhole.

"Free will" allows for us to initiate only irreversible processes with an entropy increase, not those with an entropy decrease. Thus, we cannot send an object from the future to the past. The synchronisation process of the time arrows (and the corresponding entropy growth law) forbids the initial conditions that are necessary for a macroscopic object to travel into the past (and realise the conditions for the grandfather paradox).

Paper [**46**] demonstrated that it is impossible for a thermodynamic time arrow to have the same orientation as the coordinate time arrow over a closed time-like curve due to the entropy growth law. The process of synchronisation of the time arrows described here (concerned with the infinitely large instability and ambiguity of the entropy decreasing processes) is the *physical mechanism* that actually ensures both this impossibility and the realisation of the entropy growth law over the same thermodynamic time arrow.

For microscopic wormholes, the situation is absolutely different. If the initial conditions are compatible with travel to the past through a wormhole, there are no reasons that can prevent this travel. If some small (even infinitesimally small) perturbation of the initial conditions leads to an inconsistency with the existence of the wormhole, the wormhole can always be easily destroyed **[47].** Indeed, the property of general relativity mentioned above appears; the infinitely large instability (ambiguity). This instability means that an infinitesimal perturbation of the initial conditions can result in a finite change in the final state during finite time!

However, this situation cannot be a solution to the grandfather paradox, which is a macroscopic, not a microscopic, phenomenon.

Indeed, let us suppose that there are two processes with opposite time arrow directions: a cosmonaut and the surrounding Universe. The cosmonaut travels through a wormhole from the Universe's future to the Universe's past. However, in the direction of the time arrow of the cosmonaut, the cosmonaut will be travelling from the past to the future.

In the theory of general relativity, the situation described above is impossible even in principle. Indeed, in contrast to classical mechanics, even an infinitesimal interaction leads to an infinitely large instability (ambiguity) of the process with an entropy decrease (in this case, "the process with an entropy decrease" is the cosmonaut travelling from the future to the past).

Generally, this inconsistency between macroscopic initial conditions can be accompanied both by the destruction of the wormholes [**47**] and by the conservation of the wormhole, the inversion of the cosmonaut time arrow and the erasure of his memory [**46**]**.**

In fact, *with very high probability*, the entropy growth law results in the synchronisation of the time arrows, the corresponding inversion of the cosmonaut time arrow and erasure of his memory. This law results in a *very high probability* of stability for the initially defined macroscopic space-time topology (including a set of wormholes) [**46**] and a very small probability for the destruction of macroscopic wormholes**.**

However, with very small probability, the synchronisation of time arrows can fail. This failure is a very rare large-scale fluctuation. In this case, the destruction of wormholes can occur.

In conclusion, *for macroscopic processes*, the large instability of processes with an entropy decrease, the interaction of gravity and the corresponding synchronisation of time arrows make the occurrence of initial conditions incompatible with the existence of macroscopic wormholes nearly impossible. This instability also prevents both the destruction of macroscopic wormholes and the travel of macroscopic objects to the past that results in "the grandfather paradox".

We lastly see a wonderful situation. The same reasons that allowed us to resolve the reduction paradox, and the Loschmidt and Poincare paradoxes also allow for us to resolve the information paradox for black holes and the grandfather paradox for wormholes. The universality is remarkable!

## 3. Conclusion

We described principles of new cybernetics and used these principles for resolution of basic physical paradoxes. It demonstrates universality of the principles of new cybernetics. We also see a wonderful universality for resolution of basic physical paradoxes. The same reasons that allowed us to resolve the reduction paradox, and the Loschmidt and Poincare paradoxes also allow for us to resolve the information paradox for black holes and the grandfather paradox for wormholes. The universalities are remarkable!

## Acknowledgement

This paper was written after taking part in WMSCI 2013 Conference. I would like to thank Nagib Callaos for the invitation to this very useful conference. I also would like to thank Louis H. Kauffman, Leonid I. Perlovsky, Stuart A. Umpleby, and Thomas Marlowe for very useful feedback during the conference.

**Comment on "Quantum Solution to the Arrow-of Time Dilemma" of L. Maccone.**

**Oleg Kupervasser**

Transist Video LLC
119296 Skolkovo, Russia

Scientific Research Computer Center
Moscow State University
119992 Moscow, Russia

E-Mail: olegkup@yahoo.com

*A recent letter by Maccone presents a solution based on the existing laws of quantum mechanics to the arrow-of-time dilemma. He argues that all phenomena in which the entropy decreases must not leave any information (in the observer's memory) of their having occurred because the observer is a part of the whole system. Maccone concludes that quantum mechanics is necessary to his argument, which he believes does not otherwise work in classical mechanics. This Comment consists of four parts. We discuss the basic problems in the first part. This Comment and the previously published Comment by Jennings and Rudolph describes flaws in Maccone's arguments. However, the main argument (erasure of the observer's memory), which was previously formulated in our work and was repeated by Maccone, is correct under the conditions described in this Comment. Moreover, this argument can be used to resolve a reduction paradox (the Schrödinger's Cat paradox) in quantum mechanics. This use is demonstrated in the second part. In the third part, the synchronisation (decoherence) of time arrows is discussed. In the fourth part, the synchronisation (decoherence) of time arrows in quantum gravity is considered.*

## Part 1. Macroscopic entropy, entropy increase law and memory erasure argument

### The relevance of memory erasure argument for classical mechanics.

A recent Letter by Maccone [**1**] presents a solution based on the existing laws of quantum mechanics to the arrow-of-time dilemma. He argues that all phenomena in which the entropy decreases must not leave any information (in the observer's memory) of their having occurred because the observer is a part of the whole system. He concludes that quantum mechanics (QM) is necessary to his argument, which, he believes, does not otherwise work in classical mechanics (CM). Papers [**2-4**] have clearly shown that the same arguments hold true for *both* quantum and classical mechanics. Thought experiments of both the Loschmidt (time reversal paradox) and Poincare (recurrence theorem) type are used to illustrate the arrow-of-time dilemma in the latter papers whereas Maccone uses only Loschmidt's experiment; however, he then gives a mathematical proof for the general case with entropy decrease.

The arguments to resolve both paradoxes in classical mechanics are as follows. At least in principle, CM allows exclusion of any effect from the observer on the observed system. However, most real systems are *chaotic* – a weak perturbation may lead to an exponential divergence of trajectories; also, there is always a non-negligible interaction between the observer and the observed system. Let us take the simple example of a gas expanding from a small region of space into a large volume. In this entropy-increasing process, the time evolution of the macroscopic parameters is stable to small external perturbations. After some time, if all the velocities are reversed, the gas will return to the initial small volume; this is true in the absence of any perturbation. This entropy-decreasing process is clearly unstable, and a small external perturbation would trigger continuous entropy growth. Thus, entropy-increasing processes are stable, but entropy-decreasing processes are unstable. A more rigorous theory has been developed for the general case [**2-4**]. The natural consequence of this theory is that the time arrows (whose direction is defined by the entropy growth) of both the observer and the observed system are synchronised due to the inevitable non-negligible interaction between them. Both the observer and observed system can only return to the initial state together (as the whole system) in both the Loschmidt and Poincare paradoxes; thus, the observer's memory is erased in the end. Approaching the end point, the observer's time arrow is opposite to the coordinate (absolute) time arrow, and entropy growth is observed in the entire system and in both parts of the system although the entropy decreases in the absolute time.

### The entropy increase law is FAPP law in both CM and QM

It is important to remark that the unobservability of the entropy decrease is correct only for certain *practical cases* of perturbative QM measurement experiments. For an ideal nonperturbative observation and a thermodynamically correct definition of the system entropy, the entropy decrease can *in principle* be observed in the framework of QM.

Let us first define a nonperturbative observation [**2-4**] in QM. Suppose we have some QM system in a known initial state. This initial state can be either the result of some preparation (e.g., an atom comes to the ground electronic state in vacuum after a long time) or the result of a measurement experiment (a QM system after measurement can have a well-defined state corresponding to the eigenfunction of the measured variable). We can predict further evolution of the initial wave function. Therefore, *in principle*, we can make further measurements by choosing the measured variables such that one of the eigenfunctions of the current measured variable is a current wave function of the observed system. Such a measurement process can allow for continuous observation without any perturbation of the observed quantum system. This nonperturbative observation can be easily generalised for the case of a known, *mixed* initial state.

For example, let us consider a quantum computer (QC). It has some well-defined initial state. An observer that knows this initial state can *in principle* make a nonperturbative observation of any intermediate state of the QC. However, an observer that does not know the initial state cannot make such observation because he cannot predict the intermediate state of the QC.

We can conclude that the entropy increase law is for all practical purposes (*FAPP)* law. It is correct for *perturbative* observations of macroscopic quantum systems and classical macroscopic chaotic systems due to the erasure of the observer's memory. Such small perturbations exist in any real case. However, in a general case, this is not correct.

### Correct definition of thermodynamic entropy in the Loschmidt

**paradox**

The purpose of Maccone's study was to resolve the Loschmidt paradox between the second law of thermodynamics and the reversibility of motion. Therefore, the thermodynamically correct definition of entropy, which is actually used in the formulation of the second law, must be chosen. Let us give such a definition for the entropy. Two different definitions of the entropy can be made: *macroscopic* and *ensemble* entropies [**2-4**]. The macroscopic entropy is the entropy calculated from the macroscopic parameters for *all* of the microstates that have these parameters whereas the ensemble entropy is calculated for *some* set of microstates that evolved over time from the initial state. Lastly, a standard formula for the entropy (von Neumann or classical) over the obtained distribution must be used. The second law of thermodynamics law (regarding the increase in entropy) uses the macroscopic definition of entropy.

Maccone defines the system entropy to be the sum of the ensemble entropies of the observer and the observed system ($S(A \text{ and } C) \equiv S(\rho_A) + S(\rho_C)$). For a perturbative QM observation of the macroscopic system, his definition of entropy is equivalent to that of the macroscopic entropy because both the observer and the observed system are in mixed states and correlate through microscopic variables. The classical analogues that Maccone uses to prove the necessity of QM are wrong. In his mutual entropy formula $S(A{:}C) \equiv S(\rho_A) + S(\rho_C) - S(\rho_{AC})$, the *macroscopic* entropy of the subsystems should have been used upon the transition from QM to CM, not the *ensemble* entropies. Contrary to CM, in QM both the macroscopic and ensemble entropies have the same numerical value for perturbative QM observations, although the definitions of the entropies differ. For classical macroscopic chaotic systems (the observer and the observed system) and for non-negligible interaction between the observer and observed system, the *ensemble* entropies can also be used. However, the *initial states* of the observer and observed system must be calculated from macroscopic parameters for *all* the microstates that have these parameters.

However, generally, in classical and quantum mechanics (e.g., for nonperturbative observations), this definition is not a correct definition of the thermodynamic entropy of the system. Indeed, let us consider a simple example of gas expanding from a small region of space into a large volume. This process is a macroscopic entropy-increasing process. We must use the thermodynamically correct, *macroscopic* entropy of the ideal gas: $S=kN\ln V+const$ $(T=const)$. Based on the Poincare return theorem, the gas will be very close to the initial small volume after a long time; this result is true in the absence of any perturbation. This process is a macroscopic entropy-decreasing process. In contrast to the macroscopic entropy, the ensemble entropy of the gas does not change during this evolution. Suppose we know the initial quantum state of this gas; *in principle*, we can make the nonperturbative observation described above. We therefore will be able to observe both the initial entropy increase and the final entropy decrease. This result contradicts the primary conclusion of Maccone's study. However, Maccone's considerations and conclusions are correct for the *practical case* of a *perturbative QM* observation of a macroscopic system. In this case, we used a *fixed* set of macroscopic variables for the observation. This set does not depend on the initial state (in contradiction to the nonperturbative observation described above).

We find no flaw in Maccone's *entropic considerations* within QM for perturbative observations of macroscopic systems. In contrast, D. Jennings and T. Rudolph [**5**] objected to this definition of entropy. However, the objection of D. Jennings and T. Rudolph is not relevant here [**6**] because we consider macroscopic systems. However, the examples of D. Jennings and T. Rudolph correspond to microscopic systems.

**Part 2. Schrodinger's Cat paradox and spontaneous reduction**

The complete violation of the wave superposition principle (i.e., the full vanishing of interference) and reduction of the wave function would occur only during the interaction of a quantum system with an ideal macroscopic object or device. The ideal macroscopic object either has infinite volume or consists of an infinite number of particles. Such an ideal macroscopic object can be consistently described both by quantum and by classical mechanics.

Furthermore, similarly to the classical case, we consider only systems with finite volume and a finite number of particles (unless the other is assumed). Such devices or objects can be considered to be only approximately macroscopic.

Nevertheless, a real experiment shows that even for such non-ideal macroscopic objects, the destruction of superposition and the correspondent wave function reduction may occur. We will define such a reduction for imperfect macroscopic objects as *spontaneous reduction*. Spontaneous reduction leads to paradoxes, which force one to doubt the completeness of quantum mechanics despite its tremendous successes. We will reduce the most impressive paradox from this series – the Schrodinger's Cat paradox.

Schrodinger's Cat is a thought experiment that clarifies the principle of superposition and the reduction of wave functions. A Cat is placed in a box. In addition to the Cat, there is a capsule with poisonous gas (or a bomb) in the box; this capsule (or bomb) can blow up with 50 per cent probability due to the radioactive decay of a plutonium atom or a quantum of casually illuminated light. After some time, the box is opened and one learns whether the cat is alive or not.

Until the box is opened (if the measurement is not performed), the cat remains in a very strange superposition of two states: "alive" and "dead". For macro-objects, such a situation appears very mysterious (In contrast, for quantum particles, the superposition of two different states is very natural). Nevertheless, no basic prohibition of quantum superposition for macrostates exists.

The reduction of these states upon the opening of the box by an external observer does not lead to any inconsistency with quantum mechanics. This reduction is easily explained through the interaction of the external observer with the Cat during the measurement of the Cat's state.

Nevertheless, a paradox arises for the closed box if the observer is the Cat itself. Indeed, the Cat possesses consciousness and is capable of observing both itself and the environment. Upon introspection, the Cat cannot be simultaneously alive and dead but is in just one of these two states. Experience shows that any conscious creature feels itself to be either alive or dead. Both such situations do not exist simultaneously. Therefore, the spontaneous reduction to two possible states (alive and dead) truly occurs. The Cat, even with all the contents of the box, is not an *ideal* macroscopic object. Therefore, such an observable and nonreversible spontaneous reduction contradicts reversible Schrodinger quantum dynamics. In the current case, this contradiction cannot be explained by some external influence because the system is isolated**.**

Does this system actually contradict Schrodinger quantum dynamics? When is the system macroscopic sufficient to provide the possibility for spontaneous reduction? Is it necessary for such a nearly macroscopic system to have consciousness like a Cat?

The multi-world interpretation as such does not explain the Schrodinger's Cat paradox; the interpretation only reformulates and conceals the paradox. Indeed, the Cat observes only one of the existing worlds. However, the results of further measurements depend on the correlations between the worlds. Nevertheless, neither these worlds nor these correlations are observed. "Parallel worlds" that we know nothing about can always exist. However, these worlds can truly affect the results of some future experiment of ours. That is, knowledge of the current state only (in our "world") and of the laws of quantum mechanics does not even allow us to predict the future probabilistically! However, quantum mechanics was developed for such predictions! Solely on the basis of spontaneous reduction that destroys quantum correlations between worlds, we can predict the future with knowledge of only the current (and actually observed) states of our "world". The paradox of Schrodinger's Cat returns and has only changed its shape.

Remember that the paradox of Schrodinger's Cat consists of the inconsistency between the spontaneous reduction observed by the Cat and the Schrodinger evolution that forbids such reduction. To correctly understand the paradox of Schrodinger's Cat, it is necessary to consider the paradox from the perspective of two observers: the external observer-experimenter and the Cat (i.e., *introspection*).

For the external observer-experimenter, the paradox does not arise. If the experimenter attempts to discern whether the Cat is alive or not, the experimenter inevitably influences the observable system (in agreement with quantum mechanics) and leads to the reduction. The system is not isolated and hence, cannot be described by the Schrodinger equation. The reducing role of the observer can also be played by the surrounding medium. This situation is defined as decoherence. Here, the role of the observer is more natural and is reduced to registration of the decoherence. In both cases, there is entangling of measured system with the environment or the observer, i.e., there are correlations of the measured system with the environment or the observer.

What happens if we consider a closed complete physical system that includes the observer, observed system and environment? This is the case with the Cat's introspection. The system includes the Cat and his box environment. It should be noted *that full introspection (full in the sense of quantum mechanics) and full verification of the laws of quantum mechanics are impossible in the isolated system that includes the observer*. Indeed, in principle, we can measure and analyse the state of an external system precisely. However, if we include ourselves in the consideration, there are natural restrictions. These restrictions are related to the possibility of retaining memories and analysing states of molecules with the molecules themselves. Such an assumption leads to inconsistencies. Therefore, the possibility of finding an experimental inconsistency between Schrodinger evolution and spontaneous reduction through introspection in an isolated system is also restricted.

Nevertheless, let us attempt to find some mental experiments that lead to inconsistency between Schrodinger evolution and spontaneous reduction.

1) The first example is related to the reversibility of quantum evolution. Suppose we introduce a Hamiltonian capable of reversing quantum evolution in the Cat-box system**.** Practically speaking, this process is nearly impossible; however, theoretically, no problem exists. If spontaneous reduction occurs, the process would be nonreversible. If spontaneous reduction is not present, the Cat-box system will return to an initial pure state. However, only an external observer can make such verification. The Cat cannot make it by introspection because the Cat's memory will be erased upon restoration of the initial state. From the perspective of the external observer, no paradox exists because he does not observe the spontaneous reduction that truly can lead to a paradox.
2) The second example is related to the necessity of Poincare's return of the quantum system to an initial state. Suppose the initial state was pure. If the Cat has introspection and if spontaneous reduction truly exists, it leads to a mixed state. Then return would be impossible - the mixed state cannot transfer to a pure state through the Schrodinger equation. Thus, if the Cat has fixed return, the situation is inconsistent with spontaneous reduction. However, the Cat cannot fix return (in the case of quantum mechanics fidelity) because return will erase the Cat's memory. Therefore, there is no paradox. The exterior observer actually can observe this return by measuring an initial and final state of this system. However, no paradox exists there either because the observer does not observe any spontaneous reduction that actually can lead to a paradox. It is worthwhile to note that the inconsistency between spontaneous reduction and Schrodinger evolution can be experimentally observable only if memory of the spontaneous reduction is retained by the observer and if this memory is not erased or damaged. No experiments described above are covered by this requirement. Thus, these examples clearly show that, although spontaneous reduction actually can lead to violation of Schrodinger evolution, this violation is not observed experimentally (with fidelity to quantum mechanics).
3) The third example follows: Quantum mechanics gives superposition of a live and dead Cat in a box. Theoretically, an exterior observer can always *precisely* measure this superposition if this superposition is one of the measurement eigenfunctions. Such a measurement would not destroy the superposition, in contrast to the case in which the live and dead Cat are eigenfunctions of the measurement. Having informed the Cat about the result of the measurement, we will introduce inconsistency with spontaneous reduction observed by the Cat. Such an argument has a double error. *At first*, this experiment is used for *verification* of the Cat's spontaneous reduction of existence when the observer is the Cat itself. The external observer does not influence the Cat's memory only if spontaneous reduction is not present and the Cat's state is a superposition of live and dead states. However, the observer does influence and can destroy the Cat's memory if spontaneous reduction occurs. Therefore, such an experiment cannot legitimately verify the existence of spontaneous reduction in the *past*. *Secondly*, the data transmitted to the Cat is retained in his memory. Thus, this transmission changes both the state and all further evolution of the Cat; i.e., the system cannot be considered to be isolated after the measurement. Therefore, no contradiction with the *future* exists.

The external observer does not observe spontaneous reduction and hence, does not observe the paradox. Thus, from the perspective of the external observer, verification with the aid of continuous nonperturbative observation, described in term 3, is possible and is legitimate. This verification does not influence the external observer's memory. Moreover, such verification, which does not interrupt the evolution of the observable system, allows for the measurement of not only the initial and final states of the system but also all of the intermediate states. That is, this verification implements continuous, non-perturbative observation!

It should be noted that the external observer can only theoretically observe the superposition of alive and dead Cat. Practically speaking, this observation is nearly impossible. In contrast, for small quantum systems, superposition is very observable. This difference results in the fact that quantum mechanics is generally considered to be the theory of small systems. However, for small macroscopic (mesoscopic) objects observations of superposition are also possible. A set of particles at low temperature or the states of some photons [**7**] are examples. We make an important remark**:** recently, very interesting papers were published toward the construction of mesoscopic "synergetic" systems, which are most likely similar to living organisms [**1**], [**8**], [**9**], [**10**]. It must be mentioned that the construction of such models is a problem of physics and mathematics, not philosophy.

**Part 3. Synchronisation/decoherence of time arrows.**

The follow question can arise. Let us assume that some process exists in which the entropy decreases. For definiteness, let us take this process to be the spontaneous reconstruction of a house (which was previously destroyed in an earthquake).

Let us also take the simple example of gas expanding from a small region of space into a large volume. If, after some time, all the velocities are reversed, the gas will return to the initial small volume.

If we use a camera to take a series of snapshots recording different stages of the spontaneous house construction (or gas shrinkage), we expect that the camera will record this spontaneous process. Why will the camera not be able to record it? What precisely will prevent the camera from recording these snapshots?

The answers to these questions are as follows: even a very small interaction between the camera and the observed system destroys the process of the inverse entropy decrease and results in the synchronisation of the direction of the time arrows of the observer and the observed system. (The direction of a time arrow is defined to be the direction of the entropy increase.) This very small interaction occurs because light illuminates the observed object and is reflected by the camera (and because light illuminates the camera). In the absence of the camera, the environment can act as the observer by being illuminated by and reflecting the light. (Any process without an observer is nonsensical. The observer must appear at some stage of the process; however, the influence of the observer is much smaller than the environmental influence). External noise (interaction) from the observer/the environment destroys correlation between molecules of the observed system. This noise prevents the inverse process with the entropy decrease. In quantum mechanics, such a process is defined as "decoherence". The house reconstruction (or gas shrinkage) will be stopped, i.e., the house will not actually be reconstructed/(the gas will not shrink). In contrast, processes in which the entropy increases are stable.

The following example is from Maccone's study [**1**]: "However, an observer is macroscopic by definition, and all remotely interacting macroscopic systems become correlated very rapidly (e.g., Borel famously calculated that moving a gram of material on the star Sirius by 1 m can influence the trajectories of the particles in a gas on earth on a time scale of s [**11**])"

Nevertheless, it is not a problem to reverse both the observer (the camera) and the observed system. From the Poincare return theorem for a closed system (which includes both the observer and the observed system), this return must occur automatically after a very long time. However, the memory erasure of the observer prevents this process from being registered.

The majority of real systems are *chaotic* – a weak perturbation may lead to an exponential divergence of trajectories, and there is also always a non-negligible interaction between the observed system and the observer/environment. However, *in principle*, in both quantum mechanics and classical mechanics, we can make nonperturbative observations of the entropy decrease process. A good example of such a mesoscopic device is a quantum computer: no entropy increase law exists for such a system. This device is very well isolated from the environment and the observer. However, *in practice*, nonperturbative observation is nearly impossible for macroscopic systems. We can conclude that the entropy increase law is *FAPP law*.

It should be mentioned that decoherence (synchronisation of time arrows and “entangling”) and relaxation (during relaxation, a system achieves equilibrium) are absolutely different processes. During relaxation, macroscopic variables (entropy, temperature, and pressure) change greatly to their equilibrium values, and the invisible microscopic correlations between the parts of the system increase. During decoherence, the macroscopic variables (entropy, temperature, and pressure) are nearly constant. Invisible microscopic correlations inside of the subsystems (environment, observer, and observed system) are largely destroyed; however, new correlations appear between the subsystems. This process is named “entanglement” in quantum mechanics. During this process, synchronisation of the time arrows also occurs. The relaxation time is much longer than the decoherence time.

Let us consider the synchronisation of time arrows for two systems that are non-interacting (before some initial moment). It should be mentioned that this description is made in an absolute (coordinate) system. However, both systems also have their own initially opposite time arrows, which are defined by the direction of entropy growth in each system.

This description means that there exist two non-interacting systems such that in one system time flows (i.e., entropy increases) in one direction whereas in the other system time flows in another (opposite) direction. However, if the systems come into an interaction with each other, then one system (the "stronger" one) will drag the other ("weaker") system to flow in his ("stronger") direction, so eventually, they will both have time flowing in the same direction.

“To be stronger”-what does this mean, exactly? Is this strength something that increases with the number of degrees of freedom of the system? This supposition is not correct except for small fluctuations. "Stronger" or "weaker" does not appreciably depend on the number of degrees of freedom of the systems. The interaction described above is asymmetric in the absolute (coordinate) time. For the first system, the interaction appears in its *future* after the initial moment (At the initial moment the systems have opposite time arrows) based on the time for this system. For the second system, the interaction was in its *past* based on the time of this system. Therefore, the situation is *not symmetric in time*, and the first system is always "stronger". This occurs due to the instability of processes that decrease entropy and the stability of the processes that increase entropy, as described above.

Indeed, let us consider again two initially isolated vessels of gas. In the first, the gas expands (the entropy increases). In the second, the gas shrinks (the entropy decreases).

In the first vessel, the gas expends from a small volume in the centre of a vessel. The velocities of the molecules are directed from the centre of the vessel to its boundary. It is physically clear that a small perturbation of the velocities cannot stop the expansion of gas. Indeed, after a random small perturbation, the velocities will continue to be directed from the centre of the vessel to its boundary. Noise can even increase the expansion. Therefore, the expansion process is stable.

In the second vessel, the gas shrinks from the full volume of the vessel to its centre. The velocities of all the molecules are directed toward the centre of the vessel. It is physically clear that a small random perturbation of the velocities can easily stop the shrinkage of the gas. Indeed, even after a small perturbation, the velocities will not be directed toward the centre of vessel. Thus, the shrinking process is stopped. We can therefore conclude that the shrinking process is unstable. This shrinking process can be obtained by reversing the expansion of the gas. If we reverse the velocities of the molecules of the expanding gas *before* the collisions of the molecules with each other and the vessel boundary, this instability is linear and is not strong. However, if the reversal occurs *after* these collisions, this instability is exponential and is much stronger.

Both time directions have equal roles. However, a small random noisy interaction breaks this symmetry for the two systems described above due to the instability of the entropy decrease processes. Time symmetry exists only for the *full* system that includes the two subsystems defined above. However, the time arrows of the interacting subsystems must become the same.

Instead of an interaction with infinite time $[0, +\infty]$ we can consider an interaction with a large finite time T: [0, T]. Let us choose this time T to be much smaller than the Poincare return time. Thus, in the first system we have an interaction during [0, T] based on its own time, and in the second system, the interaction is during [-T, 0] based on its own time (the “0” time moment for the first system corresponds to the “–T” time moment in the second system). Can we still apply our argument? Instead of the asymmetry of the forces, in this case, we obtain an asymmetry of the initial conditions. At the initial moment 0 for the first coordinate system [0, T], the two vessels have different time arrows. However, at the initial moment -T for the second coordinate system [-T, 0], the two vessels have the same time arrows in a negative direction.

Only if T is exactly equal to the Poincare return time will the situation indeed be symmetric. For such a situation, the two time arrows are also different at the moment T, but each arrow is opposite its initial direction at time 0. Again, the “stronger” system has the interacting forces in its future with respect to its own time arrow.

This theory can explain how entropy growth occurs in the same direction in all parts of the Universe. However, this theory cannot explain the low entropy, initial condition of the Universe. This condition is most likely a result of the anthropic principle [**12**].

## Part 4. The law of entropy increase and "synchronisation of time arrows"/decoherence in gravitational theory.

### Black Holes

In Einstein’s theory of general relativity, similarly to classical mechanics, motion is reversible. However, an important difference also exists between general relativity and classical mechanics. General relativity is an ambiguous theory. Indeed, in general relativity, two different initial states can give infinitesimally close states after a *finite* time interval. This situation occurs, for example, during the formation of a black hole because of a collapse. Let us consider the inverse process, which describes a white hole. In this process, initial states that are infinitesimally close after a *finite* time interval can give different final states. Thus, an observer/environment can considerably affect the evolution of the state during the *finite* time interval even if the observer/environment has an infinitesimally weak interaction with the white hole. We must mention that in contradiction to CM and QM, in gravitational theory, an arbitrarily small but finite interaction *always* exists. Gravity forces always exist between two arbitrary objects with nonzero masses.

Because of these two properties, the law of entropy increase is an exact law but not FAPP in general relativity theory. Therefore, entropy becomes a fundamental concept. Indeed, there is such fundamental concept as the entropy of a black hole. In addition, it is possible to explain the existence of this entropy by the perturbation created by the observer. Unlike in classical mechanics, this perturbation may now even be infinitesimally weak. During formation of a black hole, entropy increases.

Time reversal leads to the appearance of a white hole and an entropy decrease. In reality, a white hole cannot exist because of the entropy decrease. An entropy decrease is prohibited in general relativity for the same reason that it is prohibited in classical mechanics. This instability of the entropy decreasing processes is much stronger in general relativity than it is in classical mechanics. This instability results in the synchronisation of the time arrows of the white hole and of the observer/environment. The direction of the time arrow of the white hole changes to coincide with the time arrow of the observer/environment. The white hole transforms into a black hole.

Here is also the well-known black hole information paradox [**13**]**:** information (which in classical and quantum mechanics is conserved) disappears in a black hole forever. It would appear that there is no problem; the information is most likely stored inside of the black hole in some form. However, chaotic Hawking radiation makes this process of information loss explicit; the black hole evaporates, but the information is not recovered.

Hawking radiation concerns semiclassical gravitation. However, the paradox can also be formulated within the framework of the theory of general relativity. A spherical black hole can be reversed into a white hole at some moment. (This process appears impossible, but a physically similar situation with “wormholes” connecting black and white holes in different universes is considered in [**14**]). Thus, the process is converted in time. Nevertheless, information cannot be recovered due to the ambiguity (the infinitely strong instability) in the evolution of the white hole.

Usually only two solutions for this problem are considered. Either the information truly disappears or because of interior correlations of the Hawking radiation (or the exact reversal of the black hole processes after its transmutation to a white hole) the information is conserved. However, most likely, a third solution is true. Due to the inevitable influence of the observer/environment it is impossible to distinguish these two situations experimentally! However, if it is impossible to confirm this result experimentally, it is not a scientific subject.

Both in the theory of general relativity and for semiclassical gravitation the paradox can be resolved by the influence of the observer/environment. Indeed, let us suppose that Hawking radiation is correlated, not chaotic (or the white hole would be the exact inverse of the black hole). Thus, the infinitesimal influence of the observer/environment leads to the inevitable losses of these correlations (and the corresponding information) during the finite time interval. It is senseless to include the observer in the described system: complete self-description and introspection is impossible. In such a situation, the law of conservation of information cannot be confirmed experimentally even if it is actually correct.

Currently, we have no general theory of quantum gravitation. However, for a special case of a 5-dimensional anti-de-Sitter space, many scientists consider this paradox be resolved. Information is supposed to be conserved because of a hypothesis regarding AdS/CFT dualities; i.e., the hypothesis that quantum gravitation in the 5-dimensional anti-de-Sitter space (that is with a negative cosmological term) is mathematically equivalent to a conformal field theory regarding a 4-surface of this world. This hypothesis was confirmed for some special cases but is not yet proved for the general case. Suppose that this hypothesis is actually true. At first glance, this hypothesis automatically solves the information problem. Conformal field theory is unitary. If conformal field theory is actually dual to quantum gravitation, then the corresponding quantum gravitational theory is unitary as well. Therefore, in this case, information is not lost. However, we suppose that this hypothesis is not correct. The process of the formation of a black hole and its subsequent evaporation occurs on *all surfaces* of the anti-de-Sitter space (described by conformal quantum theory). This process also includes the observer/environment. However, the observer cannot precisely know the initial state and cannot analyse the behaviour of the system to verify unitarity because he is a part of this system! Hence, the observer’s influence on the system cannot be neglected. Thus, experimental verification of the information paradox again becomes impossible!

### Wormholes

Let us consider from the perspective of the entropy increase law a paradoxical object in the general relativity theory: a wormhole [**15**]**.** We will consider a Morris-Thorne wormhole [**16**]**.** Through a very simple procedure (we place one of the mouths of the wormhole on a spaceship; then the spaceship moves with relativistic velocity over a closed loop and returns the mouth to its initial location), a wormhole traversing space can be transformed into a wormhole traversing time. After this transformation, the wormhole can be used as a time machine and leads to the well-known grandfather paradox. How can this paradox be resolved?

For macroscopic wormholes, the solution can be found through the entropy increase law. The realisation of this law is ensured by the instability of entropy decreasing processes, and this instability results in the synchronisation of time arrows.

Indeed, a wormhole traversing space does not lead to a paradox. If an object enters one mouth at some moment in time, then it exits the other mouth at some later moment in time. Thus, the object travels from an initial, high-order, low-entropy environment to a future, low-order, high-entropy environment. During the trip along the wormhole, the entropy of the object also increases. Thus, the directions of the time arrows of the object and the environment are the same. The same conclusions are correct for travelling from the past to the future through a wormhole that traverses time.

However, for travel from the future to the past, the directions of the time arrows of the object and the environment are opposite. Indeed, the object travels from the initial, low-order, high-entropy environment to the high-order, low-entropy environment. However, the entropy of the object increases! As previously described, such a process is unstable and will be prevented or will be forcibly converted through a synchronisation process of the time arrows.

The initial synchronisation of a wormhole with its environment must occur when the moving mouth of the wormhole returns to its initial state during its creation.

How does the environment appear inside of the wormhole? The massive ends of the wormhole radiate. This thermodynamic radiation appears inside of the wormhole. This radiation is the environment of a traveller inside of the wormhole.

"Free will" allows for us to initiate only irreversible processes with an entropy increase, not those with an entropy decrease. Thus, we cannot send an object from the future to the past. The synchronisation process of the time arrows (and the corresponding entropy growth law) forbids the initial conditions that are necessary for a macroscopic object to travel into the past (and realise the conditions for the grandfather paradox).

Paper [**17**] demonstrated that it is impossible for a thermodynamic time arrow to have the same orientation as the coordinate time arrow over a closed time-like curve due to the entropy growth law. The process of synchronisation of the time arrows described here (concerned with the infinitely large instability and ambiguity of the entropy decreasing processes) is the *physical mechanism* that actually ensures both this impossibility and the realisation of the entropy growth law over the same thermodynamic time arrow.

For microscopic wormholes, the situation is absolutely different. If the initial conditions are compatible with travel to the past through a wormhole, there are no reasons that can prevent this travel. If some small (even infinitesimally small) perturbation of the initial conditions leads to an inconsistency with the existence of the wormhole, the wormhole can always be easily destroyed [**18**]**.** Indeed, the property of general relativity mentioned above appears; the infinitely large instability (ambiguity). This instability means that an infinitesimal perturbation of the initial conditions can result in a finite change in the final state during finite time!

However, this situation cannot be a solution to the grandfather paradox, which is a macroscopic, not a microscopic, phenomenon.

Indeed, let us suppose that there are two processes with opposite time arrow directions: a cosmonaut and the surrounding Universe. The cosmonaut travels through a wormhole from the Universe's future to the Universe's past. However, in the direction of the time arrow of the cosmonaut, the cosmonaut will be travelling from the past to the future.

In the theory of general relativity, the situation described above is impossible even in principle. Indeed, in contrast to classical mechanics, even an infinitesimal interaction leads to an infinitely large instability (ambiguity) of the process with an entropy decrease (in this case, "the process with an entropy decrease" is the cosmonaut travelling from the future to the past).

Generally, this inconsistency between macroscopic initial conditions can be accompanied both by the destruction of the wormholes [**18**] and by the conservation of the wormhole, the inversion of the cosmonaut time arrow and the erasure of his memory [**17**]**.**

In fact, *with very high probability*, the entropy growth law results in the synchronisation of the time arrows, the corresponding inversion of the cosmonaut time arrow and erasure of his memory. This law results in a *very high probability* of stability for the initially defined macroscopic space-time topology (including a set of wormholes) [**17**] and a very small probability for the destruction of macroscopic wormholes**.**

However, with very small probability, the synchronisation of time arrows can fail. This failure is a very rare large-scale fluctuation. In this case, the destruction of wormholes can occur.

In conclusion, *for macroscopic processes*, the large instability of processes with an entropy decrease, the interaction of gravity and the corresponding synchronisation of time arrows make the occurrence of initial conditions incompatible with the existence of macroscopic wormholes nearly impossible. This instability also prevents both the destruction of macroscopic wormholes and the travel of macroscopic objects to the past that results in "the grandfather paradox".

We lastly see a wonderful situation. The same reasons that allowed us to resolve the reduction paradox, and the Loschmidt and Poincare paradoxes also allow for us to resolve the information paradox for black holes and the grandfather paradox for wormholes. The universality is remarkable!